\documentclass[runningheads]{llncs}
\usepackage{graphicx}
\usepackage[hyphens]{url}
\usepackage{booktabs}
\usepackage{hyperref}
\usepackage{enumitem}
\usepackage{cite}
\usepackage{capt-of}
\usepackage{amsmath,amsfonts,bm}
\usepackage{makecell}
\usepackage{multirow}
\usepackage{placeins}
\usepackage{pifont}
\newcommand{\cmark}{\ding{51}}%
\newcommand{\xmark}{\ding{55}}%

\begin{document}
%

\title{FetMRQC: Automated Quality Control\\ for fetal brain MRI\thanks{TS is supported by the Era-net Neuron MULTIFACT project (SNSF 31NE30\_203977), OE is supported by the Swiss National Science Foundation (SNSF \#185872), the NIMH (RF1MH12186), and the CZI (EOSS5/`NiPreps'). YG acknowledges support from the SICPA foundation and EE is supported by the Instituto de Salud Carlos III (ISCIII) (AC21\_2/00016). We acknowledge access to the facilities and expertise of the CIBM Center for Biomedical Imaging, a Swiss research center of excellence founded and supported by CHUV, UNIL, EPFL, UNIGE and HUG.}}

\author{Thomas Sanchez\inst{1,2}\orcidID{0000-0003-3668-5155} \and Oscar Esteban\inst{1}\orcidID{0000-0001-8435-6191} \and Yvan Gomez\inst{3,5}\orcidID{0000-0003-0794-7750} \and Elisenda Eixarch\inst{3,4}\orcidID{0000-0001-7379-9608} \and Meritxell Bach Cuadra\inst{2,1}\orcidID{0000-0003-2730-4285}\\\email{thomas.sanchez@unil.ch}}

\authorrunning{T. Sanchez et al.}

\institute{Department of Radiology, Lausanne University Hospital (CHUV) and University of Lausanne (UNIL), Lausanne, Switzerland \and CIBM Center for Biomedical Imaging, Switzerland\and BCNatal Fetal Medicine Research Center (Hospital Clínic and Hospital Sant Joan de Déu),  Universitat de Barcelona, Spain \and IDIBAPS and CIBERER, Barcelona, Spain \and Department Woman-Mother-Child, CHUV, Lausanne, Switzerland}

\maketitle              
\begin{abstract}

Quality control (QC) has long been considered essential to guarantee the reliability of neuroimaging studies. It is particularly important for fetal brain MRI, where large and unpredictable fetal motion can lead to substantial artifacts in the acquired images. Existing methods for fetal brain quality assessment operate at the \textit{slice} level, and fail to get a comprehensive picture of the quality of an image, that can only be achieved by looking at the \textit{entire} brain volume. In this work, we propose FetMRQC, a machine learning framework for automated image quality assessment tailored to fetal brain MRI, which extracts an ensemble of quality metrics that are then used to predict experts' ratings. Based on the manual ratings of more than 1000 low-resolution stacks acquired across two different institutions, we show that, compared with existing quality metrics, FetMRQC is able to generalize out-of-domain, while being interpretable and data efficient. We also release a novel manual quality rating tool designed to facilitate and optimize quality rating of fetal brain images.

Our tool, along with all the code to generate, train and evaluate the model is available at \url{https://github.com/Medical-Image-Analysis-Laboratory/fetal_brain_qc/}

\end{abstract}
\setcounter{footnote}{0} 

\section{Introduction}
Magnetic Resonance Imaging (MRI) of the fetal brain is increasingly complementing ultrasound imaging to diagnose abnormalities, thanks to its unmatched soft tissue contrast~\cite{gholipour2014fetal,saleem2014fetal}.
However, inherent noise sources and imaging artifacts, such as fetal motion, can degrade the quality of the acquired images and jeopardize subsequent analyses.
Indeed, insufficient MRI data quality has been shown to bias neuroradiological assessment and statistical analyses~\cite{power2012spurious,reuter2015head,alexander2016subtle}.
Therefore, establishing objective image quality assessment and control (QA/QC) protocols for neuroimaging studies has long been considered critical to ensure their reliability,  generalization, and replicability~\cite{mortamet2009automatic,niso2022open}.
More recently, some QA/QC tools have been popularized for the adult brain and attempt to automate and aid exclusion decisions~\cite{esteban2017mriqc,klapwijk2019qoala,vogelbacher2019lab,samani2020qc,garcia2022brainqcnet}.
Unfortunately, these techniques have found generalization across imaging devices extremely challenging~\cite{esteban2017mriqc}.
Moreover, they are inapplicable to fetal MRI, as they rely on invalid prior knowledge, e.g., assuming that the head is surrounded by air or the relative position of the brain with respect to the axis of the bore of the scanner. This is true in particular for MRIQC~\cite{esteban2017mriqc}, from which this work is inspired. 

In fetal brain MRI, QA/QC has been approached implicitly within super-resolution reconstruction (SRR)
  methods~\cite{kuklisova-murgasova_reconstruction_2012,kainz_fast_2015,tourbier_mialsuperresolutiontoolkit_2020,uus2022retrospective,xu2023nesvor}.
SRR builds a high-resolution, isotropic 3D volume from several differently-oriented, consecutive stacks of 2D slices with substantially lower resolution on the through-plane axis (i.e., aniso\-tropic resolution).
Typically, SRR methods rely on a manual selection of stacks that are deemed adequate for reconstruction. These methods may incorporate an automated QC stage for outlier rejection that excludes sub-standard slices or pixels from the input low-resolution 
  stacks~\cite{kuklisova-murgasova_reconstruction_2012,kainz_fast_2015,ebner_automated_2020,xu2023nesvor}. Outside of fetal brain MRI, Uus et al.~\cite{uus2022automated} have explored methods to automate the rejection of poorly registered stacks. However, bad quality series can remain detrimental to the final quality of the reconstruction, even when SRR pipelines include outlier rejection schemes. This is why, in this work, we focus on the QA/QC of low-resolution (LR) stacks.
  
\begin{figure}[!t]
    \centering
    \includegraphics[width=.9\linewidth]{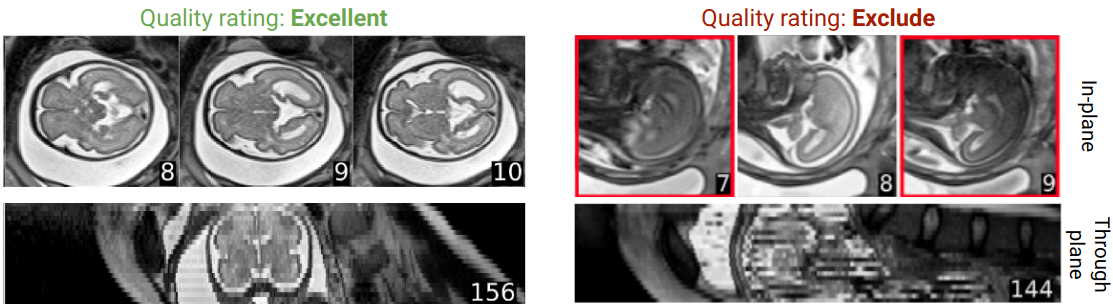}
    \caption{Examples of clinical image quality: left, excellent quality, and right, low quality. On the \textbf{right}, the stack would be excluded from further analysis due to significant intensity changes between many slices and strong signal drop; in the through-plane view, strong inter-slice motion makes it difficult to discern the structure of the brain.}
    \label{fig:qc_example_small}
\end{figure}

Recent work has applied deep learning on QA/QC of LR stacks of fetal brain MRI~\cite{lala2019deep,xu2020semi, liao2020joint}. 
These solutions automatically identify problematic slices for exclusion (QC), and, if streamlined with the acquisition, their short inference time permits re-acquiring corrupted slices~\cite{gagoski2022automated} (QA).
However, not all artifacts can be seen by analyzing single slices.
For instance, inter-slice motion (visible on the right of Figure \ref{fig:qc_example_small}), a strong bias field in the through-plane direction,
or an incomplete field of view can often clearly be seen only when considering the entire volume.

We propose FetMRQC, a framework for QA/QC of clinical fetal brain MRI, operating on stacks of LR, T2-weighted (T2w) images. 
Inspired by MRIQC~\cite{esteban2017mriqc}, FetMRQC generates a visual report corresponding to each input stack, for efficient screening and manual QA.
The tool also automatically extracts an ensemble of image quality metrics (IQMs) that reflect some quality feature.
Finally, we propose a learning framework to automatically predict image quality from the IQMs corresponding to new images.
We manually assessed more than 1000 LR stacks acquired across two different institutions and several MRI scanners within each.
Using these manual quality annotations, we quantitatively assess FetMRQC
  in two QA/QC tasks, namely, regression and binary classification (inclusion/exclusion). 
Our results demonstrate the feasibility of automated QA/QC of T2w images of the fetal brain and the
  substantial improvement of SRR after QC of subpar stacks.

\section{Methodology}
\begin{figure}[!t]
    \centering
    \includegraphics[width=0.8\linewidth]{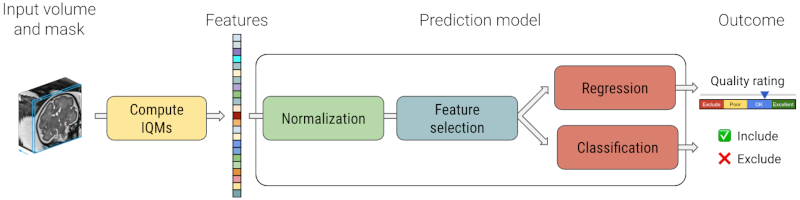}
    \caption{Proposed automated IQA pipeline. Given an input image and the corresponding brain mask, an ensemble of IQMs are extracted, and then processed in a prediction model. IQMs are normalized and irrelevant entries are removed before being fed into a regression or classification model, that produces the final outcome.}
    \label{fig:pipeline}
\end{figure}

\textit{FetMRQC} comprehends two major elements to implement QA/QC protocols of \textit{unprocessed} (LR stacks)
  fetal brain MRI data.
First, the tool builds upon MRIQC's paradigm and generates an individual QA report for each LR stack to
  assist and optimize screening and annotation by experts.
Second, \textit{FetMRQC} proposes a learning framework (Figure~\ref{fig:pipeline}) based
  on \textit{image quality metrics}~(IQMs)
  extracted from the data to automate the assessment.

\subsection{Data}
\looseness=-1
We retrospectively collected LR T2w stacks from 146 subjects retrieved from two existing databases at two different institutions, CHUV and BCNatal. The data were acquired on different Siemens scanners and a common MR scheme (Half fourier Single-shot Turbo spin-Echo; HASTE), at 1.5T and 3T, with both normal and pathological cases. The full list of parameters is detailed in Table~\ref{tab:scanners} in the Appendix. 
The corresponding local ethics committees independently approved the studies under which data were collected, and all participants gave written informed consent. 
CHUV provided 61 subjects, with an average of 7.9$\pm$3.0 LR stacks per subject. The 
BCNatal provided 85 subjects, $5.8\pm3.4$ stacks per subject.
The aggregate sample size is $N$=1010 LR series.

\subsection{Manual QA of fetal MRI stacks}

\begin{figure}[!t]
    \centering
    \includegraphics[width=.65\linewidth]{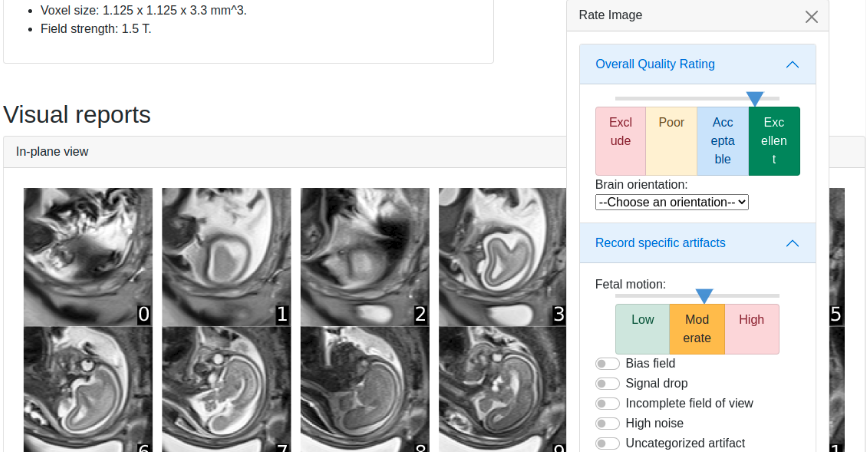}
    \caption{Visual quality assessment tool proposed in this work. The HTML report gives some general information on the anonymized scan, displays all slices with brain content, as well as both orthogonal through-plane views, as seen on Figure~\ref{fig:qc_example_small}.}
    \label{fig:qc_report}
\end{figure}

Akin to MRIQC~\cite{esteban2017mriqc}, \textit{FetMRQC} generates an HTML-based report adapted to the QA of fetal brains
  for each input LR stack (Figure \ref{fig:qc_report}).
The input dataset is required to comply with the Brain Imaging Data Structure (BIDS)\cite{gorgolewski2016brain},
  a format widely adopted in the neuroimaging community.
The reports are generated using an image with a corresponding brain mask.
This mask can be extracted automatically, and in this work, we used MONAIfbs~\cite{ranzini2021monaifbs}.
Each individual-stack report has a QA utility (the so-called \textit{rating widget}, with which raters can fill in an overall quality score,
  the in-plane orientation, and the presence and grading of artifacts showcased by the stack.
We chose also to use an interval (as opposed to categorical) rating scale with four main quality ranges:
  $[0,1]$: exclude -- $(1,2]:$ poor -- $(2,3]$: acceptable -- $(3,4]$: excellent.
Interval ratings simplify statistical modeling, set lower bounds to \textit{annotation noise}, and 
  enable the \textit{inference} task where a continuous quality score is assigned to input images,
  rather than broad categories.
  
We generated the individual report corresponding to each of the 1010 LR stacks.
Two raters independently annotated each 555 of the dataset using the proposed tool, 
  which yielded 100 stacks annotated by both raters to test inter-rater variability.
These 100 stacks were randomly selected.
Rater~, YG, 1 is a maternal-fetal physician with 5 years of experience in neuroimaging, 
  and Rater~2, MBC, is an engineer with 20 years of experience in neuroimaging.

\subsection{IQMs extraction and learning}

\subsubsection{IQMs in fetal brain MRI.}\label{sec:iqm}
Leveraging the same workflow that generates the individual reports, a number of IQMs were extracted from the
  full stack and from the information within the brain mask. 
Since only a few IQMs defined by MRIQC can be applied to fetal brains, we implemented a set of IQMs specific to the application at hand.
In~\cite{kainz_fast_2015}, Kainz et al. proposed a \texttt{rank\_error} based on the estimated compressibility of an LR stack.
Later on, brain mask-derived measures were used such as the volume of the brain mask, \texttt{mask\_volume}, in~\cite{ebner_automated_2020} or its \texttt{centroid} in ~\cite{dedumast2020translating}. Recently, deep learning-based IQM slice-wise and stack-wise have been proposed, \texttt{dl\_slice\_iqa}\cite{xu2020semi} and \texttt{dl\_stack\_iqa}~\cite{legoretta2020DL}, which respectively do slice-wise and stack wise quality assessment\footnote{We use the \textit{pre-trained} models throughout these experiments, as we want to test the off-the-shelf value of these IQMs}\footnote{The method of Liao et al.~\cite{liao2020joint} was not included because their code is not publicly available, and we could not get in contact with the authors.}.

We propose additional IQMs for quality prediction that have not previously been used in the context of fetal brain MRI. They can roughly be categorized into two groups, and full details are provided in Table~\ref{tab:metrics} in the Appendix. In a nutshell, \textbf{intensity-based} IQMs directly rely on the voxel values of the image. These include summary statistics~\cite{esteban2017mriqc} like mean, median, and percentiles. We also re-use metrics traditionally used for outlier rejection~\cite{kuklisova-murgasova_reconstruction_2012,kainz_fast_2015,ebner_automated_2020} to quantify the intensity difference between slices in a volume. We also compute entropy~\cite{esteban2017mriqc}, estimate the level of bias using N4 bias field correction~\cite{tustison_n4itk_2010} and estimate the sharpness of the image with Laplace and Sobel filters. The second type of metrics are \textbf{shape-based} and operate directly on the automatically extracted brain mask. We propose to use a morphological closing to detect through-plane motion, as well as edge detection to estimate the variation of the surface of the brain mask. 

\textbf{Variation on the metrics.} All the IQMs operate on LR images or masks and can be modified by various transformations of the data. For instance, Kainz et al.~\cite{kainz_fast_2015} evaluated their metrics only on the third of the slices closest to the center of a given LR volume. We include various variations on the IQMs, including considering the whole ROI instead of the centermost slices (\texttt{\_full}). Other variations include keeping or masking the maternal background, aggregating point estimates using mean, median, or other estimators, computing information theoretic metrics on the union or intersection of masks, etc. Finally,  \texttt{slice\_loss} metrics can be either computed as a pairwise comparison between all slices (by default) or only on a window of neighboring slices (\texttt{\_window}). With all the different variations, we arrive at a total of $75$ different features. See Fig.~\ref{fig:crosscorr} in the Appendix for a cross-correlation matrix between all IQMs on our dataset.

\textbf{Feature-wise normalization.} We explored both global and subject-wise standard and robust scaling. In our experiments, we did not notice substantial changes between global and subject-wise scaling, and for the sake of simplicity, we adopted global scaling in the rest of this work.

\textbf{Feature-selection and dimensionality reduction.} We explored feature variance and correlations. We removed irrelevant features ($\text{variance}=0$ for a given subset of data) and drop IQMs that are highly correlated with each other (with threshold $0.8$ and $0.9$). We also removed features which do not contribute more than noise using the Winnow algorithm~\cite{littlestone1988learning} with extremely randomized trees~\cite{esteban2017mriqc}. Finally, we also explored using principal component analysis to reduce dimensionality and construct orthogonal features as input of the model.

\textbf{Model selection and evaluation through nested cross-validation.}
Nested cross-validation (CV) is a fully automated mechanism to perform model selection and evaluation
  without introducing optimistic biases~\cite{varoquaux2017assessing}.
In this framework, an outer CV loop benchmarks the expected performance of the family of models represented
  by the inner loop.
An inner CV loop is executed for each fold of the outer loop to select the best-performing model on average.

We set up the nested CV framework with five folds in both the inner and outer loops
  ($80\%$ train, $20\%$ test at each level), for both the regression and classification tasks.
We ensured to group together all the LR stacks of each subject to avoid data leaking between the training and testing fold.
For the regression task, we evaluated linear regression, gradient boosting, and random forests.
For classification, we considered logistic regression, random forests, gradient boosting, and AdaBoost.
We primarily optimized feature selection strategies and model fitting parameters.
A detailed description of the combinations of models and parameters optimized is available in Table \ref{tab:params} in the Appendix, and selected parameters are detailed in Table~\ref{tab:nested_cv}.
The experiments were implemented with Python 3.9.15 and Scikit-learn 1.1.3~\cite{scikit-learn}.

\noindent\textbf{Baseline models and evaluation.}
We evaluated three variants of our model. First of all, we assessed the predictive power of a subset of individual features 
  that have previously been proposed and used for fetal brain QA/QC
  (namely, \texttt{rank\_error}, \texttt{rank\_error\_full} \texttt{mask\_volume}, \texttt{centroid}, \texttt{centroid\_full},\linebreak \texttt{dl\_slice\_iqa} and \texttt{dl\_stack\_iqa}).
The individual features were scaled and then fitted with a linear regression or logistic regression model for prediction.
We then reported the performance of the best-performing feature, which consistently was the \texttt{dl\_stack\_iqa}~\cite{legoretta2020DL}.
Secondly, we considered a \textsc{base} variant of our IQA pipeline, using the same subset of features above.
Finally, we considered a model using all available features, referred to as \textsc{FetMRQC}. 

The models were evaluated on two different data configurations. In the \textit{in-domain} evaluation, we aimed at quantifying how models would generalize on new subjects acquired at either sites. 
We used nested CV to tune the hyperparameters on the data from one group of subjects from both sites and evaluated the model on a different group of subjects.
In the \textit{out-of-domain} evaluation, we aimed at quantifying how models would generalize to new sites. We used nested CV to tune the hyperparameters on the data from one site and evaluated the model on the other site.

Our regression results were evaluated using mean absolute error (MAE) and Spearman rank correlation.
Our classification results used F1-score and the area under the receiver operating characteristic curve (AUC ROC).
\section{Results and discussion}
\subsubsection{\textit{FetMRQC}'s reports optimize stack screening and bolster inter-rater reliability.}\label{exp:rating}
\begin{figure}[!t]
  \begin{minipage}[t]{0.42\textwidth}
    \centering
    \vspace{0pt}\includegraphics[width=.92\linewidth]{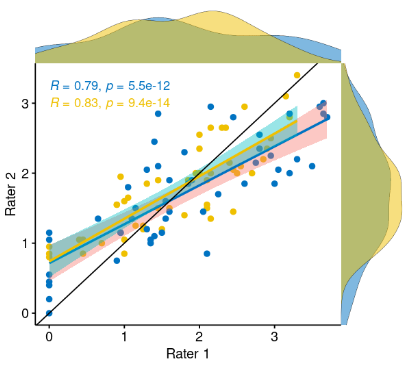}
    \caption{Inter-rater correlation: blue and yellow data are respectively from CHUV and BCNatal.} 
      \label{fig:corr_plot}
  \end{minipage}
  \hfill
  \begin{minipage}[t]{0.57\textwidth}
    \centering
    \vspace{0pt}\includegraphics[width=.8\linewidth]{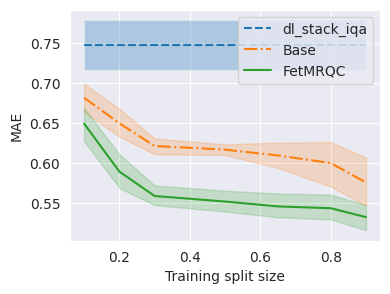}
    \caption{MAE as a function of the size of the training set. The individual metric, \texttt{dl\_stack\_iqa\_full} is constant because it only performs inference.}
      \label{fig:regr_tr_size}
  \end{minipage}
\end{figure}

Figure \ref{fig:corr_plot} summarizes the high agreement between the two raters, with a Pearson correlation coefficient of $0.79$ and $0.83$ for CHUV and BCNatal. 
On CHUV, 120 series are manually rated below the exclude threshold $\text{Quality} <1$), and 414 above. On BCNatal, 151 series are rated excluded, and 393 rated above. The total include-to-exclude ratio is 2.98. The total rating time was around 5h 40min for Rater 1 (median of 37s per volume), and around 6h10 for Rater 2 (median of 40s per volume).

\begin{figure}[!t]
  \begin{minipage}[c]{0.49\textwidth}
    \centering
    \vspace{0pt}\captionof{table}{In-domain evaluation}\label{tab:id_perf}
    \resizebox{\linewidth}{!}{
    \begin{tabular}{lcccc}
    \toprule
    & \multicolumn{2}{c}{Regression} & \multicolumn{2}{c}{Classification}\\
    \cmidrule(lr){2-3} \cmidrule(lr){4-5}
    & MAE ($\downarrow$) & Spearman ($\uparrow$) & F1 ($\uparrow$)  & AUC ($\uparrow$)\\
    \midrule   
    \texttt{dl\_stack\_iqa}  &$0.72\pm0.05$ &$0.37\pm0.11$ &$0.85\pm0.02$ &$0.53\pm0.02$\\
    Base        &$0.58\pm0.05$ &$0.60\pm0.03$ &$0.88\pm0.02$ &$0.71\pm0.06$\\
    FetMRQC        &$\mathbf{0.53\pm0.09}$ &$\mathbf{0.71\pm0.05}$&$\mathbf{0.90\pm0.02}$ & $\mathbf{0.77\pm0.07}$\\
    \bottomrule
    \end{tabular}}
  \end{minipage}
  \hfill
  \begin{minipage}[c]{0.49\textwidth}
    \vspace{0pt}\captionof{table}{Out of domain evaluation}\label{tab:ood_perf}
    \resizebox{\linewidth}{!}{
    \begin{tabular}{lccccc}
    \toprule
    & \multicolumn{2}{c}{Regression} & \multicolumn{2}{c}{Classification}\\
    \cmidrule(lr){2-3} \cmidrule(lr){4-5}
    & MAE ($\downarrow$) & Spearman ($\uparrow$) & F1 ($\uparrow$) & AUC ($\uparrow$)\\
    \midrule   
    \texttt{dl\_stack\_iqa}   &$0.75\pm0.03$ &$0.42\pm0.10$ & $0.83\pm0.04$&$0.57\pm0.07$\\
    Base        &$0.75\pm0.01$ &$0.38\pm0.03$ & $0.85\pm0.02$&$0.67\pm0.06$\\
    FetMRQC        &$\mathbf{0.68\pm0.06}$ &$\mathbf{0.50\pm0.08}$& $\mathbf{0.89\pm0.01}$& $\mathbf{0.77\pm0.05}$\\
    \bottomrule
    \end{tabular}}
  \end{minipage}
  
  \begin{minipage}[t]{0.49\textwidth}
    \centering
    \vspace{0pt}\includegraphics[width=\linewidth]{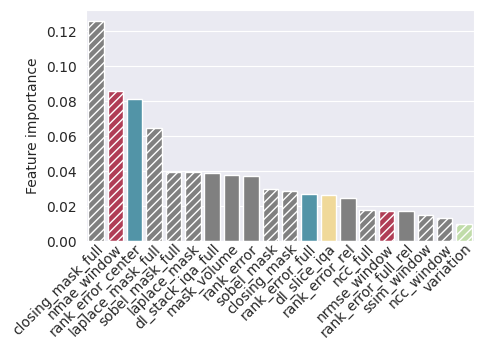}
  \end{minipage}
  \begin{minipage}[t]{0.49\textwidth}
    \centering
    \vspace{0pt}\includegraphics[width=1.\linewidth]{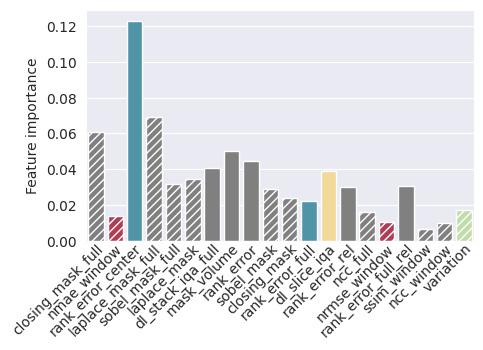}
  \end{minipage}
  \vspace{-.3cm}
   \caption{Feature importance averaged on regression (left) and classification models (right), for the $20$ largest contributors. Gray features are not correlated with each other ($|\text{corr}| < 0.75$). Features with the same color are correlated ($>0.75$) with each other. Hatched features are features that were proposed in this work.}
      \label{fig:feat_imp}

    \includegraphics[width=\linewidth]{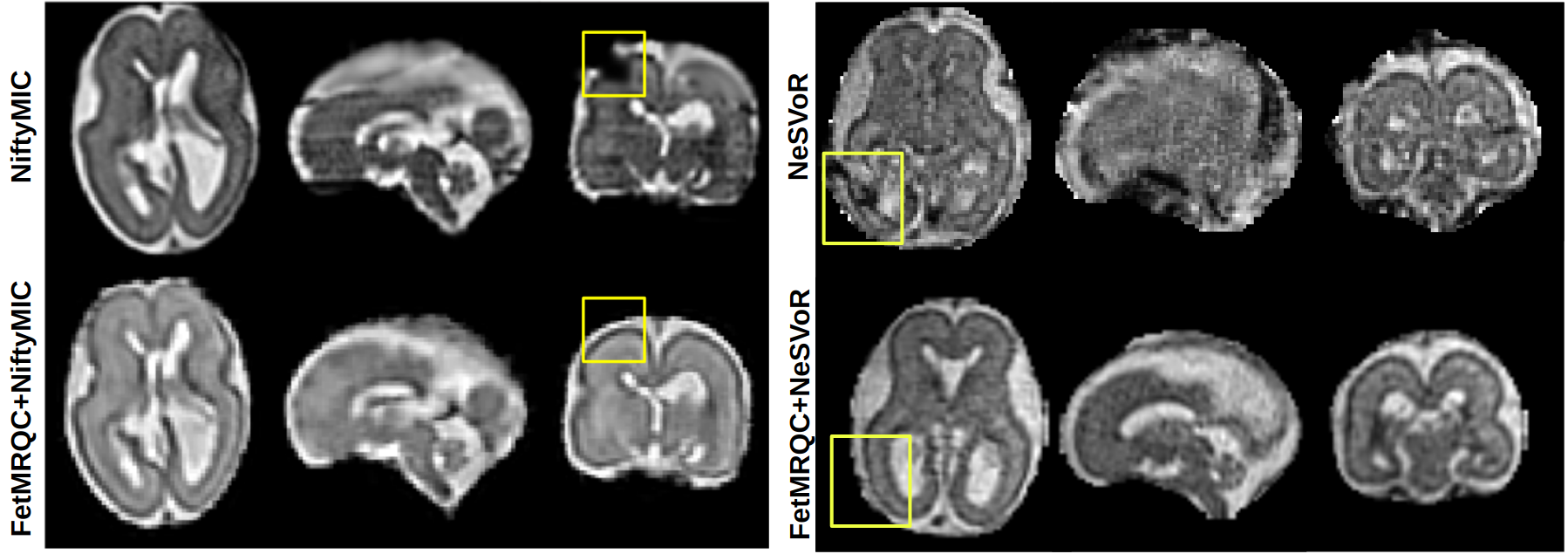}
    \vspace{-.5cm}
    \caption{Illustration of usefulness of QC on the SR reconstruction, using NiftyMIC~\cite{ebner_automated_2020}, and NeSVoR~\cite{xu2023nesvor}. FetMRQC+NiftyMIC successfully removes 6 series out of the 13 available, and FetMRQC+NeSVoR removes 2 out of 5. FetMRQC leads to a substantially greater image quality.}\label{fig:qc_sr}
    \vspace{-.4cm}
\end{figure}

\noindent\textbf{New IQMs drive \textit{FetMRQC}'s performance.}
 Tables \ref{tab:id_perf} and \ref{tab:ood_perf} shows that the additional features substantially increase the performance of the models, both in the case of regression and classification, and both in- and out-of-domain. 

\noindent\textbf{Using more IQMs allows to generalize with fewer training data within domain.}
On Figure~\ref{fig:regr_tr_size}, we see that not only using more features reduces the overall error on the validation set, but allows to reach a good testing performance using as little as $30\%$ ($\sim 40$ subjects, $300$ LR series) of the data for training.

\noindent\textbf{\textit{FetMRQC}'s classifier is robust when trained on one site and evaluated on the other.} 
On Table \ref{tab:ood_perf}, we see that classification is retained when training on one of the sites and evaluating on the other. Although regression performance is hurt by the domain gap between sizes, our model still generalizes best.

While \texttt{dl\_stack\_iqa} was trained on different data than FetMRQC, the out-of-domain evaluation provides a fair evaluation, where each method is evaluated on a different dataset than the one used for training it (see the acquisition parameters in Table~\ref{tab:scanners}).

\noindent\textbf{Different features matter for different tasks.}
An advantage of feature-based methods over deep learning-based methods is that their decisions are based on a combination of simple features, which make them more interpretable. On Figure \ref{fig:feat_imp}, we see a comparison of the feature importance for regression and classification tasks. These were obtained as by averaging the feature importance for the model selected at each fold of the nested CV. We observe that the importance of features varies largely with the task at hand. While the closing of the brain mask (\texttt{closing\_mask\_full}) and the nMAE are the most predictive of the regression quality), the exclusion of slices relies more on the compressibility of the centermost slices of the image (\texttt{rank\_error\_center}) and the edges of the brain mask (\texttt{laplace\_mask\_full}). 

\noindent\textbf{\textit{FetMRQC}'s classifier can drastically improve the outcome of SRR methods.}
Figure \ref{fig:qc_sr} shows how bad quality stacks can drastically impact the final quality of the SRR, using NiftyMIC~\cite{ebner_automated_2020} and NeSVoR~\cite{xu2023nesvor}.
The images reconstructed without QC displayed significant artifacts. Conversely, the two SRR methods showcased improved quality when subpar stacks were excluded. 

\section{Conclusion}
We propose \textit{FetMRQC}, which adapts the approach of a popular tool, MRIQC, to implement reliable
  QA/QC of low-resolution stacks of T2w MRI of the fetal brain.
Using \textit{FetMRQC} tools, two experts assessed the quality of 1010 stacks with high inter-rater
  reliability; and these annotations were then applied to automated regression (for QA) and classification (QC).
Nested cross-validation of a set of models and hyper-parameters showed how QA and QC can be
  automated.
Objective and reliable QA/QC procedures are critical to ensure the reliability and repeatability
  of neuroimaging studies, and \textit{FetMRQC} demonstrates how existing approaches can readily
  be applied on fetal brain MRI.
  
\bibliographystyle{IEEEtran}
\bibliography{biblio}
\newpage
\appendix
\section*{Appendix}
\begin{table}[!b]
\centering
    \caption{Detailed description of the metrics proposed for FetMRQC.}
    \vspace{-0.3cm}
    \resizebox{.9\linewidth}{!}{\begin{tabular}{m{2.3cm}m{12.5cm}}
    \toprule
    \multicolumn{2}{l}{\textsc{Intensity-based metrics}}\\
    \midrule
    \texttt{slice\_loss} & Use metrics commonly used for outlier rejection~\cite{kuklisova-murgasova_reconstruction_2012,kainz_fast_2015,ebner_automated_2020} to compute the difference between slices in the volume. We consider (normalized) mean averaged error, (normalized) mutual information, normalized cross correlation, (normalized) root mean squared error, peak signal-to-noise ration, structural similarity and joint entropy.\\
    \texttt{sstats}~\cite{esteban2017mriqc}     & Compute the mean, median, standard deviation, percentiles $5\%$ and $95\%$, coefficient of variation and kurtosis on brain ROI.\\
     \texttt{entropy}~\cite{esteban2017mriqc} & Measure the overall entropy of the image.\\
    \texttt{bias} & Level of bias estimated using N4 bias field correction~\cite{tustison_n4itk_2010} \\
    \texttt{filter\_image} & Estimate the sharpness by using Laplace and Sobel filters (commonly used for edge detection)\\
    \midrule
    \multicolumn{2}{l}{\textsc{Shaped-based metrics}}\\
    \midrule
    \texttt{closing\_mask} & Morphological closing of the brain mask in the through-plane direction, to detect inter-slice motion. Report the average difference with the original mask.\\
    \texttt{filter\_mask}& Estimate the sharpness of the brain mask using Laplace and Sobel filtering. In an ideal case, the brain mask would be smoothly varying, especially in the through-plane direction.\\
    \bottomrule
    \end{tabular}}
    \label{tab:metrics}
    \vspace{.5cm}
    \caption{Detailed description of the data from CHUV and BCNatal. Field refers to the magnetic field of the scanner, TR is the repetition time and TE is the echo time and FoV is the Field of View. All scanners used a  Half-Fourier Acquisition Single-shot Turbo spin Echo imaging (HASTE) sequence. }\label{tab:scanners}
    \vspace{-0.3cm}
    \resizebox{.8\linewidth}{!}{\begin{tabular}{lclcccc}
    \toprule
    \multicolumn{6}{c}{\textsc{\textbf{CHUV}}}\\
    \textbf{Model (Siemens)} & Field [T] & $(n_{\text{subjects}},n_{\text{LR}})$ & TR [ms] & TE [ms] & Resolution [$\text{mm}^3$]  & FoV [cm]\\
    \midrule
    \textbf{Aera}            &1.5    &$(34,281)$ &  1200 & 90    & $1.12\times1.12\times3.3$ & $35.8$\\
    \textbf{MAGNETOM Sola}   & 1.5   &$(17, 138)$& 1200  & 90    & $1.1\times1.1\times3.3$ & $35.8$\\
    \textbf{MAGNETOM Vida}~~~& 3     &$(2, 14)$  & 1100  & 101   & $0.55\times0.55\times3$ & $35.2$\\
    \textbf{Skyra}           &3      &$(8, 77)$  &  1100 & 90    & $0.55\times0.55\times3$ & $35.2$\\
    \textbf{Avanto}          &1.5    &$(1, 5)$   & 1000  & 82    & $1.2\times1.2\times4$ & $30.0$\\
    \midrule
    \multicolumn{6}{c}{\textsc{\textbf{BCNatal}}}\\
    \textbf{Model (Siemens)} & Field [T] & $(n_{\text{subjects}},n_{\text{LR}})$ & TR [ms] & TE [ms] & Resolution [$mm^3$]  & FoV [cm]\\
    \midrule
    \textbf{Aera}   & 1.5   & $\mathbf{(16, 158)}$ &&&\\
                            &       &-~$(6,80)$  & 1500& 82& $0.55\times0.55\times2.5$ & $28.2$\\
                            &       &-~$(4,34)$   & 1000& 137& $0.59\times0.59\times3.5$ & $22.7$ / $30.2$\\
                            &       &-~$(4, 33)$ & 1000&81 & $0.55\times0.55\times3.15$ & $28.2$\\
                            &       &-~$(2,11)$   & 1200& 94& $1.72\times1.72\times4.2$ & $35.8$/$44.0$\\
    \textbf{MAGNETOM Vida}   &   3   & $(11, 56)$& 1540 & 77 & $1.04\times1.04\times3$ & $20.0$\\
    \textbf{TrioTim}         &3      & $\mathbf{(59, 322)}$&&&  \multicolumn{2}{r}{\textit{// 4 outliers}}\\
                            &       &-~$(24,97)$& 1100&127&$0.51\times0.51\times3.5$ & $26.1$\\
                            &       & -~$(15,108)$& 990& 137& $0.68\times0.68\times3.5-6.0$ &$26.1$\\
                            &       &-~$(14,71)$& 2009& 137& $0.51\times0.51\times3.5$ &$26.1$\\
                            &       &-~$(1,14)$ & 3640& 137& $0.51\times 0.51\times 3.5$ &$26.1$\\

    \bottomrule
    \end{tabular}
    }
    \vspace{.5cm}\captionof{table}{Parameters automatically optimized by the inner loop of the nested CV.}
    \vspace{-0.3cm}
    \label{tab:params}
    \resizebox{.55\linewidth}{!}{
    \begin{tabular}{lc}
    \toprule
    \thead{Model step} & \thead{Parameters}\\
    \midrule
    Remove correlated features & $\text{Threshold} \in \{0.8,0.9\}$; Disabled  \\
    Data Scaling & \makecell{Standard scaling, Robust scaling,\\ No scaling}  \\
    Winnow algorithm & Enabled, Disabled\\
     PCA & Enabled, Disabled\\
    Regression models & \makecell{Linear regression, Gradient\\ boosting, Random Forest} \\
    Classification models & \makecell{Logistic regression, Random Forest,\\ Gradient Boosting, AdaBoost}\\
    \bottomrule
    \end{tabular}}
    \end{table}
    
\FloatBarrier

\begin{figure}[!ht]
\centering
    \captionof{table}{Selected hyperparameters for the different nested cross validation procedures. The in-domain experiment uses 5-fold nested cross-validation, while the out-of-domain experiment splits data by site (CHUV and BCNatal) and as a result has only two folds. The list of possible parameters is provided in Table~\ref{tab:params}.}\label{tab:nested_cv}
    \resizebox{.496\linewidth}{!}{\begin{tabular}{lccccc}
    \toprule
    \multicolumn{6}{c}{In-domain -- Regression}\\
    \midrule
    &Remove feat. & Scaling & Winnow & PCA & Model \\
    \multirow{5}{*}{\textsc{Base}} 
    & \xmark & Standard& \cmark& \cmark& Random Forest\\
    & 0.9 & Standard& \cmark& \xmark& Random Forest\\
    & 0.8 & Standard& \cmark& \xmark& Random Forest\\
    & 0.8 & Standard& \cmark& \xmark& Gradient Boosting\\
    & 0.8 & Standard& \cmark& \xmark& Random Forest\\[2mm]

    \multirow{5}{*}{\textsc{FetMRQC}} 
    & \xmark& Robust& \cmark& \cmark& Random Forest\\
    & \xmark& Robust& \cmark& \cmark& Gradient Boosting\\
    & 0.8& None& \cmark& \xmark& Random Forest\\
    & 0.9& Robust& \cmark& \xmark& Gradient Boosting\\
    & \xmark& None& \cmark& \xmark& Random Forest\\
    \midrule
    \multicolumn{6}{c}{Out-of-domain -- Regression}\\
    \midrule
    &Remove feat. & Scaling & Winnow & PCA & Model \\
    \multirow{2}{*}{\textsc{Base}} 
    & 0.9& Robust& \cmark & \xmark & Gradient Boosting \\
    & \xmark& None& \cmark& \cmark& Random Forest \\[2mm]
    \multirow{2}{*}{\textsc{FetMRQC}} 
    & 0.9& None& \cmark& \cmark& Lin. Regression\\
    & \xmark& Robust& \cmark& \xmark& Random Forest\\
    \bottomrule
    \end{tabular}}
    \hfill
    \resizebox{.496\linewidth}{!}{\begin{tabular}{lccccc}
     \toprule
    \multicolumn{6}{c}{In-domain -- Classification}\\
    \midrule
    &Remove feat. & Scaling & Winnow & PCA & Model \\
    \multirow{5}{*}{\textsc{Base}} 
    & 0.8& Robust   & \cmark& \cmark& Random Forest\\
    & \xmark& None     & \cmark& \cmark& Random Forest\\
    & \xmark& Standard & \cmark& \xmark& Random Forest\\
    & 0.8& Standard & \cmark& \xmark& Random Forest\\
    & \xmark& Standard & \cmark& \xmark& Random Forest\\[2mm]
    \multirow{5}{*}{\textsc{FetMRQC}} 
    & \xmark& Standard & \cmark& \cmark& Gradient Boosting\\
    & 0.9& Robust   & \cmark& \cmark& Random Forest\\
    & 0.8& Robust   & \cmark& \xmark& Gradient Boosting\\
    & 0.9& Standard & \cmark& \cmark& Lin. Regression\\
    & \xmark& Standard & \cmark& \xmark& Gradient Boosting\\
    \midrule
    \multicolumn{6}{c}{Out-of-domain -- Classification}\\
    \midrule
    &Remove feat. & Scaling & Winnow & PCA & Model \\
    \multirow{2}{*}{\textsc{Base}} 
    & 0.8& Standard& \cmark&\xmark & Random Forest\\
    & 0.8& Standard& \cmark&\cmark & Random Forest\\[2mm]
    \multirow{2}{*}{\textsc{FetMRQC}} 
    & 0.9& Standard& \cmark&\xmark & Random Forest\\
    & 0.8& Standard& \cmark&\xmark & Random Forest\\
    \bottomrule
    \end{tabular}}
    
    \vspace{.4cm}
    \includegraphics[width=\linewidth]{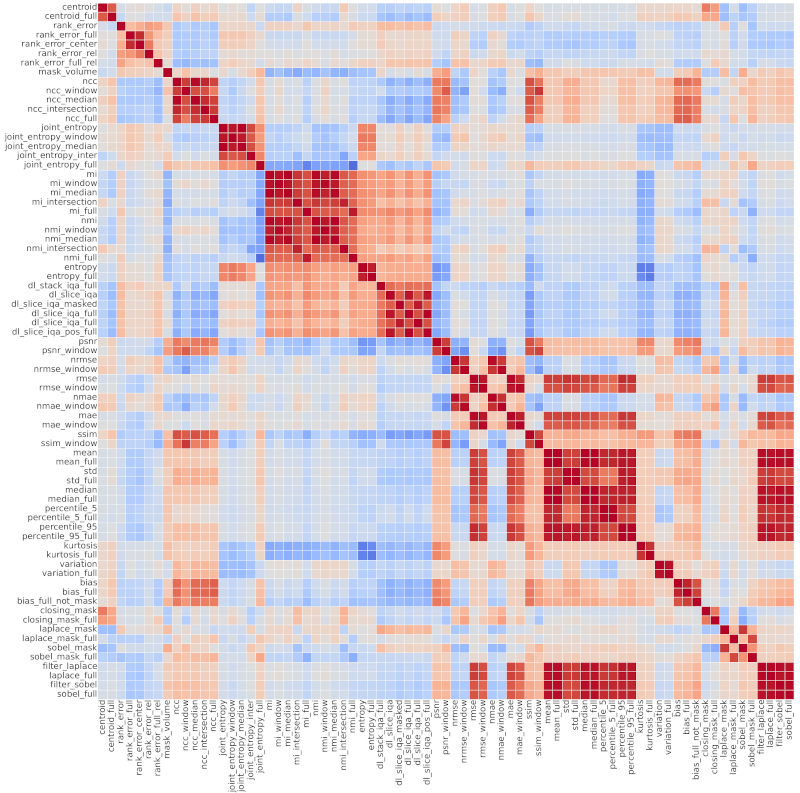}
    \caption{Correlation matrix between all 75 IQMs, evaluated on the entire dataset. Blue refers to negative correlations, and red to positive ones.}
    \label{fig:crosscorr}
\end{figure}

\end{document}